\documentclass[prd,twocolumn,floatfix,superscriptaddress,showpacs,letterpaper,altaffilletter,nofootinbib]{revtex4} 
\usepackage{color} 
\usepackage{times} 
\usepackage{graphicx} 
\usepackage{fancyhdr} 
\usepackage{float} 
\usepackage{ulem} 
\makeatletter 
\makeatletter 
\def\@fnsymbol#1{\ifcase#1\or * \or  $+$ \or  \$ \or \#  \or \dag \or \ddag \or 
$\mathsection$ \or $ \mathparagraph$ \or $\|$  \or \textordfeminine \or \textbullet    
\or ** \or $++$ \or  \$\$ \or \#\#  \or \dag\dag \or \ddag\ddag \or 
$\mathsection\mathsection$ \or $ \mathparagraph\mathparagraph$ \or $\|\|$  \or  
\textordfeminine\textordfeminine \or \textbullet \textbullet \or *** \or $+++$  
\or  \$\$\$ \or \#\#  \or \dag\dag \or \ddag\ddag \or 
$\mathsection \mathsection\mathsection$ \or $ \mathparagraph  
\mathparagraph\mathparagraph$ \or $\|\|\|$  \or  
\textordfeminine\textordfeminine\textordfeminine \or  
\textbullet\textbullet\textbullet \or \else \@ctrerr\fi} 
\makeatother 
 
\newcommand\chisqthresh{\ensuremath{\xi^\ast}} 
\newcommand{\upperlimit}{49} 
 
\def\thercsid{\relax} 
\def\rcsid#1{\def\next##1#1{\def\thercsid{##1}}\next} 
\rcsid$Id: ligo-tama-inspiral.tex,v 1.38 2006/03/20 21:31:57 sfairhur Exp $ 
 
\renewcommand{\today}{\number\day\space\ifcase\month\or 
  January\or February\or March\or April\or May\or June\or 
  July\or August\or September\or October\or November\or December\fi 
  \space\number\year} 
 
\begin{document} 
 
\title{Joint LIGO and TAMA300 Search for Gravitational Waves from Inspiralling
Neutron Star Binaries}


%


\newcommand*{\AG}{Albert-Einstein-Institut, Max-Planck-Institut f\"ur Gravitationsphysik, D-14476 Golm, Germany}
\affiliation{\AG}
\newcommand*{\AH}{Albert-Einstein-Institut, Max-Planck-Institut f\"ur Gravitationsphysik, D-30167 Hannover, Germany}
\affiliation{\AH}
\newcommand*{\AN}{Australian National University, Canberra, 0200, Australia}
\affiliation{\AN}
\newcommand*{\CH}{California Institute of Technology, Pasadena, CA  91125, USA}
\affiliation{\CH}
\newcommand*{\DO}{California State University Dominguez Hills, Carson, CA  90747, USA}
\affiliation{\DO}
\newcommand*{\CA}{Caltech-CaRT, Pasadena, CA  91125, USA}
\affiliation{\CA}
\newcommand*{\CU}{Cardiff University, Cardiff, CF2 3YB, United Kingdom}
\affiliation{\CU}
\newcommand*{\CL}{Carleton College, Northfield, MN  55057, USA}
\affiliation{\CL}
\newcommand*{\CO}{Columbia University, New York, NY  10027, USA}
\affiliation{\CO}
\newcommand*{\TDAMSUT}{Department of Advanced Materials Science, The University of Tokyo,  Kashiwa, Chiba 277-8561, Japan} 
\affiliation{\TDAMSUT}
\newcommand*{\TDAUT}{Department of Astronomy, The University of Tokyo,  Bunkyo-ku, Tokyo 113-0033, Japan} 
\affiliation{\TDAUT}
\newcommand*{\TDPHU}{Department of Physics, Hiroshima University,  Higashi-Hiroshima, Hiroshima 739-8526, Japan} 
\affiliation{\TDPHU}
\newcommand*{\TDUIU}{Department of Physics, University of Illinois at Urbana-Champaign, Urbana, IL 61801, USA} 
\affiliation{\TDUIU}
\newcommand*{\TDPMUE}{Department of Physics, Miyagi University of Education,  Aoba Aramaki, Sendai 980-0845, Japan} 
\affiliation{\TDPMUE}
\newcommand*{\TDPUT}{Department of Physics, The University of Tokyo,  Bunkyo-ku, Tokyo 113-0033, Japan} 
\affiliation{\TDPUT}
\newcommand*{\TERIUT}{Earthquake Research Institute, The University of Tokyo,  Bunkyo-ku, Tokyo 113-0032, Japan} 
\affiliation{\TERIUT}
\newcommand*{\ER}{Embry-Riddle Aeronautical University, Prescott, AZ   86301 USA}
\affiliation{\ER}
\newcommand*{\TFSKU}{Faculty of Science, Kyoto University,  Sakyo-ku, Kyoto 606-8502, Japan} 
\affiliation{\TFSKU}
\newcommand*{\TFSNU}{Faculty of Science, Niigata University,  Niigata, Niigata 950-2102, Japan} 
\affiliation{\TFSNU}
\newcommand*{\TFSTHU}{Faculty of Science and Technology, Hirosaki University,  Hirosaki, Aomori 036-8561, Japan} 
\affiliation{\TFSTHU}
\newcommand*{\TGSASUT}{Graduate School of Arts and Sciences, The University of Tokyo,  Meguro-ku, Tokyo 153-8902, Japan} 
\affiliation{\TGSASUT}
\newcommand*{\TGSSOCU}{Graduate School of Science, Osaka City University,  Sumiyoshi-ku, Osaka 558-8585, Japan} 
\affiliation{\TGSSOCU}
\newcommand*{\TGSSOU}{Graduate School of Science, Osaka University,  Toyonaka, Osaka 560-0043, Japan} 
\affiliation{\TGSSOU}
\newcommand*{\TGSSTU}{Graduate School of Science, Tohoku University,  Sendai, Miyagi 980-8578, Japan} 
\affiliation{\TGSSTU}
\newcommand*{\THEARO}{High Energy Accelerator Research Organization,  Tsukuba, Ibaraki 305-0801, Japan} 
\affiliation{\THEARO}
\newcommand*{\HC}{Hobart and William Smith Colleges, Geneva, NY  14456, USA}
\affiliation{\HC}
\newcommand*{\TICRRUT}{Institute for Cosmic Ray Research, The University of Tokyo,  Kashiwa, Chiba 277-8582, Japan} 
\affiliation{\TICRRUT}
\newcommand*{\TILSUEC}{Institute for Laser Science, University of Electro-Communications,  Chofugaoka, Chofu, Tokyo 182-8585, Japan} 
\affiliation{\TILSUEC}
\newcommand*{\IU}{Inter-University Centre for Astronomy  and Astrophysics, Pune - 411007, India}
\affiliation{\IU}
\newcommand*{\CT}{LIGO - California Institute of Technology, Pasadena, CA  91125, USA}
\affiliation{\CT}
\newcommand*{\LM}{LIGO - Massachusetts Institute of Technology, Cambridge, MA 02139, USA}
\affiliation{\LM}
\newcommand*{\LO}{LIGO Hanford Observatory, Richland, WA  99352, USA}
\affiliation{\LO}
\newcommand*{\LV}{LIGO Livingston Observatory, Livingston, LA  70754, USA}
\affiliation{\LV}
\newcommand*{\LU}{Louisiana State University, Baton Rouge, LA  70803, USA}
\affiliation{\LU}
\newcommand*{\LE}{Louisiana Tech University, Ruston, LA  71272, USA}
\affiliation{\LE}
\newcommand*{\LL}{Loyola University, New Orleans, LA 70118, USA}
\affiliation{\LL}
\newcommand*{\MP}{Max Planck Institut f\"ur Quantenoptik, D-85748, Garching, Germany}
\affiliation{\MP}
\newcommand*{\MS}{Moscow State University, Moscow, 119992, Russia}
\affiliation{\MS}
\newcommand*{\ND}{NASA/Goddard Space Flight Center, Greenbelt, MD  20771, USA}
\affiliation{\ND}
\newcommand*{\NA}{National Astronomical Observatory of Japan, Tokyo  181-8588, Japan}
\affiliation{\NA}
\newcommand*{\TNIAIST}{National Institute of Advanced Industrial Science and Technology,  Tsukuba, Ibaraki 305-8563, Japan} 
\affiliation{\TNIAIST}
\newcommand*{\NO}{Northwestern University, Evanston, IL  60208, USA}
\affiliation{\NO}
\newcommand*{\TOU}{Ochanomizu University,  Bunkyo-ku, Tokyo 112-8610, Japan} 
\affiliation{\TOU}
\newcommand*{\TPEDTU}{Precision Engineering Division, Faculty of Engineering, Tokai University,  Hiratsuka, Kanagawa 259-1292, Japan} 
\affiliation{\TPEDTU}
\newcommand*{\TRESCEU}{Research Center for the Early Universe (RESCEU), Graduate School of Science, The University of Tokyo, Tokyo 113-0033, Japan}
\affiliation{\TRESCEU}
\newcommand*{\TRIKEN}{RIKEN,  Wako, Saitaka 351-0198, Japan} 
\affiliation{\TRIKEN}
\newcommand*{\SC}{Salish Kootenai College, Pablo, MT  59855, USA}
\affiliation{\SC}
\newcommand*{\SE}{Southeastern Louisiana University, Hammond, LA  70402, USA}
\affiliation{\SE}
\newcommand*{\SA}{Stanford University, Stanford, CA  94305, USA}
\affiliation{\SA}
\newcommand*{\SR}{Syracuse University, Syracuse, NY  13244, USA}
\affiliation{\SR}
\newcommand*{\PU}{The Pennsylvania State University, University Park, PA  16802, USA}
\affiliation{\PU}
\newcommand*{\TC}{The University of Texas at Brownsville and Texas Southmost College, Brownsville, TX  78520, USA}
\affiliation{\TC}
\newcommand*{\TTDU}{Tokyo Denki University,  Chiyoda-ku, Tokyo 101-8457, Japan} 
\affiliation{\TTDU}
\newcommand*{\TTIT}{Tokyo Institute of Technology,  Meguro-ku, Tokyo 152-8551, Japan} 
\affiliation{\TTIT}
\newcommand*{\TR}{Trinity University, San Antonio, TX  78212, USA}
\affiliation{\TR}
\newcommand*{\HU}{Universit{\"a}t Hannover, D-30167 Hannover, Germany}
\affiliation{\HU}
\newcommand*{\BB}{Universitat de les Illes Balears, E-07122 Palma de Mallorca, Spain}
\affiliation{\BB}
\newcommand*{\BR}{University of Birmingham, Birmingham, B15 2TT, United Kingdom}
\affiliation{\BR}
\newcommand*{\FA}{University of Florida, Gainesville, FL  32611, USA}
\affiliation{\FA}
\newcommand*{\GU}{University of Glasgow, Glasgow, G12 8QQ, United Kingdom}
\affiliation{\GU}
\newcommand*{\MD}{University of Maryland, College Park, MD 20742 USA}
\affiliation{\MD}
\newcommand*{\MU}{University of Michigan, Ann Arbor, MI  48109, USA}
\affiliation{\MU}
\newcommand*{\OU}{University of Oregon, Eugene, OR  97403, USA}
\affiliation{\OU}
\newcommand*{\RO}{University of Rochester, Rochester, NY  14627, USA}
\affiliation{\RO}
\newcommand*{\UW}{University of Wisconsin-Milwaukee, Milwaukee, WI  53201, USA}
\affiliation{\UW}
\newcommand*{\VC}{Vassar College, Poughkeepsie, NY 12604}
\affiliation{\VC}
\newcommand*{\TWU}{Waseda University,  Shinjyuku-ku, Tokyo 169-8555, Japan} 
\affiliation{\TWU}
\newcommand*{\WU}{Washington State University, Pullman, WA 99164, USA}
\affiliation{\WU}
\newcommand*{\TYITPKU}{Yukawa Institute for Theoretical Physics, Kyoto University,  Sakyo-ku, Kyoto 606-8502, Japan} 
\affiliation{\TYITPKU}



\author{B.~Abbott}    \affiliation{\CT}
\author{R.~Abbott}    \affiliation{\CT}
\author{R.~Adhikari}    \affiliation{\CT}
\author{A.~Ageev}    \affiliation{\MS}  \affiliation{\SR}
\author{J.~Agresti}    \affiliation{\CT}
\author{P.~Ajith}    \affiliation{\AH}
\author{B.~Allen}    \affiliation{\UW}
\author{J.~Allen}    \affiliation{\LM}
\author{R.~Amin}    \affiliation{\LU}
\author{S.~B.~Anderson}    \affiliation{\CT}
% ...

 \collaboration{The LIGO Scientific Collaboration, http://www.ligo.org}
 \noaffiliation


\author{T.~Akutsu} \affiliation{\TICRRUT}
\author{T.~Akutsu} \affiliation{\TDAUT}
\author{M.~Ando} \affiliation{\TDPUT}
\author{K.~Arai} \affiliation{\NA}
\author{A.~Araya} \affiliation{\TERIUT}
\author{H.~Asada} \affiliation{\TFSTHU}
\author{Y.~Aso} \affiliation{\TDPUT}
\author{P.~Beyersdorf} \affiliation{\NA}
\author{Y.~Fujiki} \affiliation{\TFSNU}
\author{M.-K.~Fujimoto} \affiliation{\NA} 
% ...

 \collaboration{The TAMA Collaboration}
 \noaffiliation

\date{\today}

\begin{abstract}  
 
We search for coincident gravitational wave signals from inspiralling neutron
star binaries using LIGO and TAMA300 data taken during early 2003.  Using a
simple trigger exchange method, we perform an inter-collaboration coincidence
search during times when TAMA300 and only one of the LIGO sites were
operational.  We find no evidence of any gravitational wave signals.  We place
an observational upper limit on the rate of binary neutron star coalescence
with component masses between  1 and 3$M_{\odot}$ of $\upperlimit$ per year
per Milky Way equivalent galaxy at a $90\%$ confidence level.  The methods
developed during this search will find application in future network inspiral
analyses.
 
\end{abstract} 
 
 
\maketitle 
 
The first generation of gravitational wave interferometric detectors are
rapidly approaching their design sensitivities.  These include the
LIGO~\cite{ref:ligo} and TAMA300~\cite{ref:tama} detectors as well as
GEO~\cite{ref:geo} and Virgo~\cite{ref:virgo}. Inspiralling binaries of
neutron stars and/or black holes are one of the most promising sources of
gravitational radiation for these detectors.  Indeed, several searches for
such signals have already been completed~\cite{ref:LIGO_S1_ins, TAMAdt2,
LIGOS2iul, TAMADT8iul}.  In the long term, the chances of detecting
gravitational waves from a binary inspiral are greatly improved by making
optimal use of data from all available detectors.  The immediate benefit of a
multi-detector coincidence search is a significant reduction in the the false
alarm rate for a fixed detection efficiency.  Additionally, a search involving
all available detectors will provide an increase in observation time when, for
example, at least two detectors are operating.  The different orientations of
the detectors make them sensitive to different parts of the sky, thus a
combined search can lead to improved sky coverage.  If an event is detected in
multiple instruments it is possible to localize the position of the source and
improve parameter estimation.  In addition, independent observations in
well-separated detectors using different hardware and analysis algorithms
would increase confidence in a detection, while reducing the possibility of an
error or bias.  

The importance of joint searches has long been acknowledged, and indeed
several network searches have previously been completed.  A network of
resonant decectors was used to carry out a joint search for gravitational wave
bursts \cite{igec}, and more recently data from the LIGO and TAMA300 detectors
were used to perform a joint burst search \cite{ltburst}.  In this paper, we
present the first inter-collaboration search for gravitational waves from the
binary inspiral of neutron stars using modern large scale interferometric
detectors. This represents an important step towards a global network analysis
of gravitational wave data.  Furthermore, this search provides a firm basis
for development of network analysis techniques.  
 
The joint coincidence search described here uses data from the second LIGO
science run (S2) which occurred at the same time as the eighth TAMA300 data
taking run (DT8) in 2003.  The LIGO S2 data have already been searched for
gravitational waves from binary neutron stars~\cite{LIGOS2iul}.  That search
used only data in which both of the LIGO sites were operational.  In this
paper, we report on a coincidence search using LIGO and TAMA300 data when only
one LIGO site was operating in coincidence with the TAMA300 detector.  The
LIGO data analyzed in this paper was not analyzed in Ref.~\cite{LIGOS2iul}.
During S2 and DT8, the LIGO detectors were an order of magnitude more
sensitive than TAMA300.  However, since TAMA300 was sensitive to the majority
of candidate sources in the Milky Way, a joint coincidence search provides
information about inspiraling neutron star binaries in the galaxy.  Further,
by performing this joint search between the LIGO and TAMA collaborations, we
are able to significantly increase in the length of time searched in
coincidence during the S2/DT8 run.  Since LIGO and TAMA300 were the only large
interferometers which were operated during S2/DT8 period, it is important to
perform a joint analysis.  

The data from each of the detectors are searched independently for event
candidates, or ``triggers'' \cite{igec}.  The details of these triggers, such
as the coalescence time and the masses of the component stars, are then
exchanged between collaboration members, and the triggers are searched for
coincidences.  The coincidence requirements of the search are determined by
adding simulated signals to the data streams of the detectors, and determining
the accuracy with which various parameters are
recovered~\cite{LIGO-TAMA_gwdaw}.  The exchange of single instrument triggers
and subsequent coincidence analysis is quite simple and does not involve the
exchange of large amounts of interferometer data.  It provides a natural first
step in an inter-collaboration analysis.  If an interesting candidate event
were found, it would then be followed up by an optimal, fully coherent
analysis of the data around the time of the candidate.  In this joint
LIGO--TAMA300 search, we find no evidence of any inspiral signals in the data
and so we place an observational upper limit on the rate of binary neutron
star coalescence in the Milky Way galaxy.  

The LIGO network of detectors consists of a 4km interferometer ``L1'' in
Livingston, LA and a 4km ``H1'' and a 2km ``H2'' interferometer which share a
common vacuum system in Hanford, WA. TAMA300 is a 300m interferometer ``T1''
in Mitaka, Tokyo.  Basic information on the position and orientation of these
detectors and detailed descriptions of their operation can be found in
Refs.~\cite{ref:ligo, ref:tama}.  The data analyzed in this search was taken
during LIGO S2, TAMA300 DT8 between 16:00 UTC 14 February 2003 and 16:00 UTC
14 April 2003.  We only analyze data from the periods when both LIGO and
TAMA300 interferometers were operating.  Furthermore, we restrict to times
when only one of the LIGO sites was operational.  Therefore, we have four
independent data sets to analyze: the data set during which neither H1 nor H2
were operating --- the nH1-nH2-L1-T1 coincident data set (here ``n'' stands
for ``not operating'') --- and three data sets when one or both of the Hanford
detectors were operational but L1 was not ---  the H1-H2-nL1-T1,
H1-nH2-nL1-T1, and nH1-H2-nL1-T1 coincident data sets.  

\begin{figure} 
\begin{center} 
\includegraphics[width=2.0in]{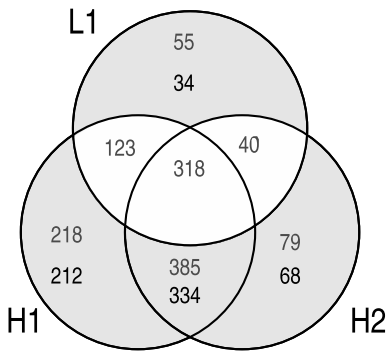} 
\end{center} 
\caption{The number of hours that each combination of detectors was searched 
during the S2/DT8 run.  The upper number gives the amount of time the specific 
LIGO detectors were coincidentally operational.  The lower number gives the 
total amount of time searched in coincidence with TAMA300.  The shaded region 
corresponds to the data used in this search.} 
\label{fig:ligotama_venn} 
\end{figure}

During the S2 science run, a strong correlation was found in the L1
interferometer between inspiral triggers and non-stationary noise in the
auxiliary channel, L1:LSC-POB\_I, which is proportional to the length
fluctuations of the power recycling cavity.  Therefore, we apply a veto to
exclude times of excess noise in POB\_I, details of which are given in Ref.
\cite{LIGOS2iul}.  No efficient veto channels were found for the H1, H2 or T1
detectors.  After applying the veto to L1, there are 34 hours of nH1-nH2-L1-T1
data.  Additionally, there are 334 hours of H1-H2-nL1-T1 data, 212 hours of
H1-nH2-nL1-T1 data and 68 hours of nH1-H2-nL1-T1 data, giving a total
observation time of 648 hours.  The data used in this search are summarized in
Figure \ref{fig:ligotama_venn}.  

To avoid any bias from tuning our pipeline using the same data from which we
derive our upper limits, the tuning of analysis parameters was done without
examining the full coincident trigger sets.  Instead, parameter tuning was
done on the \textit{playground} data which consists of approximately 10\% of
the data chosen as a representative sample.  In this analysis, the length of
playground data is 64 hours.  The analysis of the playground data and tuning
of the search is described in more detail in Ref.  \cite{LIGO-TAMA_gwdaw}.
The playground data \textit{was} searched for candidate gravitational wave
events, but was excluded from the data set used to place the upper limit.
Subtracting the playground data leaves a total of 584 hours of non-playground
data used in placing the upper limit. 
 
In a search for inspiralling neutron star binaries, we can characterize the
sensitivity of the detectors by their maximum observable effective distance,
or range.  This is defined as the distance at which an inspiral of $1.4$--$1.4
M_{\odot}$ neutron stars, in the optimal direction and orientation with
respect to each detector, would produce a signal to noise ratio (SNR) of 8.
The effective distance of a signal is always greater than or equal to the
actual distance.  On average it is $2.3$ times as large as the actual
distance, with the exact factor dependent upon the source location and
orientation relative to the detector.  During the S2 science run the ranges of
the LIGO detectors, averaged over the course of the run, were 2.0, 0.9 and 0.6
Mpc for L1, H1 and H2 respectively.  This made them sensitive to signals from
the Milky Way and favorably oriented potential sources in the local group of
galaxies.  The range of TAMA300 during DT8 was 52 kpc, making it sensitive to
the majority of the Milky Way.  Thus, the detectors were sensitive to a
similar population of candidate sources.  Since we require a signal to be
observed in both the LIGO and TAMA300 detectors, for this search we restrict
our attention to gravitational waves produced by inspiralling neutron star
binaries in the Milky Way.
 
The search methods employed in this paper are similar to those used in the
LIGO S2 search~\cite{LIGOS2iul} and the independent TAMA300 DT8
search~\cite{TAMAdt8}.  Therefore, in this paper we will not describe the LIGO
or TAMA300 analysis pipelines in great detail, but instead emphasize the
differences between this search and those described previously.   
 
For the LIGO search, we split the data into analysis blocks of 2048 seconds
length, overlapped by 128 seconds.  For each block, we construct a template
bank with a minimal match of $97\%$ and component masses between 1 and
3$M_{\odot}$~\cite{ref:Owen_and_Sathya}. We analyze the data using the
\textsc{findchirp} implementation of matched filtering for inspiral signals in
the LIGO Algorithm Library~\cite{Allen:2005fk,LALS2LT}.  The most important
thresholds used in the LIGO search are given in Table~\ref{tab:ligo_thresh}.
Most notably, we use an SNR threshold $\rho^\ast = 7$ for matched filtering.
Additionally, we perform a waveform consistency ($\chi^{2}$)
test~\cite{Allen:2004}.  For this, we require the power observed in the signal
to be evenly distributed between $p$ frequency bands.  The threshold is 
\begin{equation} \chi^{2} \le (p + \delta \, \rho^{2}) \xi^{\ast} \, .
\label{eq:chisq} \end{equation} 
We use a higher threshold on SNR ($7$ rather than $6$) and also a tighter
$\chi^{2}$ threshold ($5$ rather than $12.5$ in the Hanford detectors) than in
the LIGO only S2 inspiral analysis.  This is due to the fact that we limit our
attention to signals from the Milky Way which tend to have a large SNR in the
LIGO S2 data stream.  The tighter thresholds vastly reduce the false alarm
rate while giving a negligible loss of detection efficiency.  

For times during which both the H1 and H2 detectors were operational, we
perform a triggered analysis of H2, as described in detail in
Ref.~\cite{LIGOS2iul}.  We produce a template bank and matched filter the H1
data.  Only for those times and masses that we obtain a trigger in H1 do we
filter the H2 data.  This significantly reduces our analysis time while having
no effect on the detection efficiency.  We then search for triggers coincident
in time and mass between the H1 and H2 detectors.  The use of a triggered
search allows us to require the mass parameters of coincident triggers to be
identical.  Studies performed by injecting simulated signals show we can
determine the end time of an inspiral to within $1 \mathrm{ms}$ and
consequently we use this as our time coincidence window.  Finally, we
implement an amplitude consistency test between triggers in H1 and
H2~\cite{LIGOS2iul}; this includes keeping any triggers from H1 whose
recovered effective distance renders them unobservable in the less sensitive
H2 detector. 
 
\begin{table} 
\begin{tabular}{clr} 
Parameter & Description & value \\ 
\hline  
MM & Templatebank Minimal Match & 97\% \\ 
$\rho^\ast$ & Matched Filter Threshold & 7.0 \\ 
$p$ & Number of $\chi^{2}$ bins & 15 \\ 
$\delta$ & $\chi^{2}$ threshold parameter & 0.023 \\ 
$\chisqthresh$ & $\chi^{2}$ threshold parameter& 5.0 \\ 
$\delta t_{HH}$ & H1/H2 Timing Coincidence & 1.0 ms \\ 
$\delta m_{HH}$ & H1/H2 Mass Coincidence & 0 \\ 
\end{tabular} 
\caption{A list of the most significant parameters used for the search of 
the LIGO data.} 
\label{tab:ligo_thresh} 
\end{table} 
 
For the TAMA300 search,  we split the data into analysis blocks of 52.4288
seconds length. The adjacent blocks of data are overlapped by 4.0 seconds in
order not to lose signals which lie on the border of two adjacent blocks.  We
construct a template bank with a minimal-match of
$97\%$~\cite{ref:Tanaka_and_Tagoshi} for each locked segment, in which the
detector was continuously operated without any interruptions.  The most
significant thresholds in the TAMA300 search are listed in
Table~\ref{tab:tama_thresh}.  We use a SNR threshold $\rho^\ast = 7$ for
matched filtering.  In the TAMA300 only search, we introduce a threshold on
the value of $\rho/\sqrt{\chi^2}$ to reduce the number of false alarms
\cite{TAMAdt8, ref:tama-lism}.  However, in the LIGO--TAMA300 analysis, we
introduce a $\chi^2$ threshold as in Eq.~(\ref{eq:chisq}).  By cutting on
$\chi^{2}$, the number of triggers is significantly reduced.  In addition,
some of the coincidence analysis becomes much simpler since LIGO and TAMA300
use a similar criterion for $\chi^2$.  More details of the TAMA300 analysis
pipeline are available in Ref.~\cite{TAMAdt8, ref:tama-lism}.  
 
\begin{table} 
\begin{tabular}{clr} 
Parameter & Description & value \\ 
\hline  
MM & Templatebank Minimal Match & 97\% \\ 
$\rho^\ast$ & Matched Filter Threshold & 7.0 \\ 
$p$ & Number of $\chi^{2}$ bins & 16 \\ 
$\delta$ & $\chi^{2}$ threshold parameter & 0.046 \\ 
$\chisqthresh$ & $\chi^{2}$ threshold paramater& 2.3 \\ 
\end{tabular} 
\caption{A list of the most significant parameters used for the search of 
the TAMA300 data.} 
\label{tab:tama_thresh} 
\end{table} 
 
The requirements for coincidence between triggers in the LIGO and TAMA300 
detectors are determined by adding simulated inspiral events to the data 
streams of the detectors.  Thresholds are chosen so that injected signals seen 
separately in both the LIGO and TAMA300 detectors survive the coincidence step 
with near 100\% efficiency, while minimizing the rate of accidental 
coincidences.  Since both the LIGO and TAMA300 pipelines can accurately 
determine the coalescence time and mass of an injected signal, it is natural 
to require consistency of these values in our coincidence test.  We measure 
the accuracy with which these parameters are recovered in each detector and 
set the coincidence window to be the sum of these accuracies.  The values of 
time and mass coincidence parameters are given in 
Table~\ref{tab:coinc_thresh}.  Both pipelines recover the end time with an 
accuracy of 1 ms, to which we must add the light travel time between sites to 
obtain the values given in the table.  The mass parameter most accurately 
recovered by the pipelines is the chirp mass of a signal.  The chirp mass is 
defined as $\mathcal{M} = M \eta^{3/5}$, where $M = m_{1} + m_{2}$ is the 
total mass of the system and $\eta = m_{1} m_{2} / M^{2}$ is the dimensionless 
mass ratio.  To pass coincidence, we require the chirp masses of two triggers 
to agree within $0.05 M_{\odot}$.  Further details of how these parameters were 
chosen are available in Ref.~\cite{LIGO-TAMA_gwdaw}.  
 
\begin{table} 
\begin{tabular}{clr} 
Parameter & Description & value \\ 
\hline  
$\delta t_{HT}$ & Timing between Hanford and TAMA & 27.0 ms \\ 
$\delta t_{LT}$ & Timing between Livingston and TAMA & 35.0 ms \\ 
$\delta \mathcal{M}$ & Chirp mass window & 0.05 $M_{\odot}$\\ 
\end{tabular} 
\caption{The coincidence windows used for the LIGO--TAMA300 search.} 
\label{tab:coinc_thresh} 
\end{table} 
 
\begin{figure} 
\begin{center} 
\includegraphics[width=\linewidth]{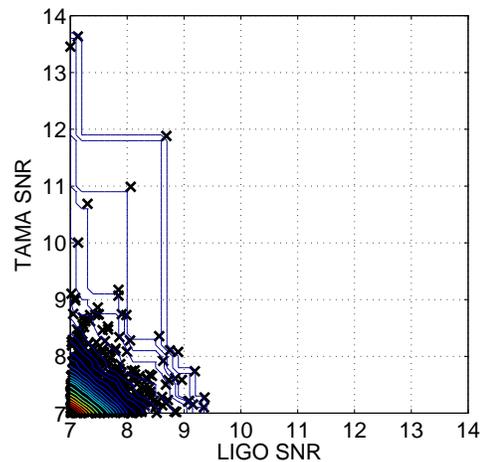} 
\end{center} 
\caption{The signal to noise ratios $\rho_{\mbox{\rm\tiny LIGO}}$ vs 
$\rho_{\mbox{\rm\tiny TAMA}}$ of the accidental coincident triggers using 
100 time-shifts.  The contours of constant false alarm probability are also  
shown.} 
\label{fig:shift_trigs} 
\end{figure}

The coincidence parameters described above were chosen to provide a good 
efficiency to simulated events.  However, there is some chance that noise 
induced events in the detectors might survive our coincidence tests.  In order 
to estimate the background of such chance coincident triggers we perform a 
time-shift analysis~\cite{ref:Amaldi_Astone}.  To do this, we time-shift the 
TAMA300 triggers by multiples of 5 seconds and search for coincidence between 
the time-shifted TAMA300 triggers and LIGO triggers.  We perform 100 time-shifts,
with a value of the time-shift ranging from $-250$ to $250$ seconds. 
These shifts are much longer than the light travel time between the sites, so 
that any coincidence cannot be from actual gravitational waves.  They are also 
longer than the typical detector noise auto-correlation time, longer than the 
longest signal template duration (4 seconds) and shorter than typical 
timescales of detectors' non-stationarity, so that each time-shift provides an 
independent estimate of the accidental coincident rate.  The SNRs of the 
triggers obtained from the time-shift analysis are plotted in 
Figure~\ref{fig:shift_trigs}.  The plot shows that the distribution of 
background coincidences does not follow the circular false alarm contours 
expected for Gaussian noise~\cite{ref:Pai}.  Instead, a statistic which more 
accurately reflects the constant false alarm probability contours is the sum 
of the SNR in the two detectors,  
\begin{equation}\label{eq:rho_combined} 
  \rho_{\mbox{\rm\tiny C}} = \rho_{\mbox{\rm\tiny LIGO}}  
    + \rho_{\mbox{\rm\tiny TAMA}}\, .  
\end{equation} 
We use this statistic in our analysis to distinguish background triggers 
from detection candidates.   
 
\begin{figure} 
\begin{center} 
\includegraphics[width=\linewidth]{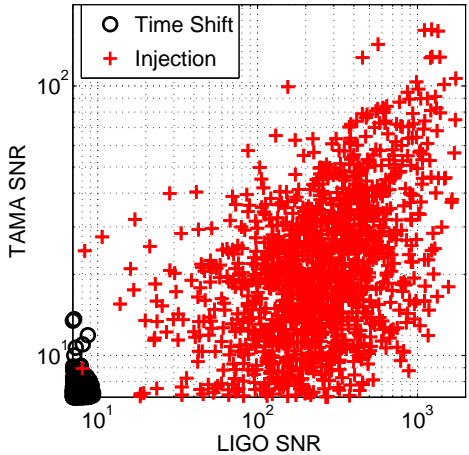} 
\end{center} 
\caption{The signal to noise ratios $\rho_{\mbox{\rm\tiny LIGO}}$ vs 
$\rho_{\mbox{\rm\tiny TAMA}}$ of the triggers associated with injections ($+$) and 
those from accidental coincidences arising in 100 time-shifts ($\circ$).}  
\label{fig:shift_inj_trigs} 
\end{figure}

To measure the sensitivity of the search, we perform a set of injections into
both sets of data.  The simulated waveforms added to the data consist of
galactic binary neutron star inspiral signals.  The majority of neutron stars
in the Milky Way lie in the galactic bulge, which we take to have a radius of
4 kpc and height of 1.5 kpc.  The sun is assumed to lie 8.5 kpc from the
center of the galaxy.  Further details of the galactic model used are
available in Ref.~\cite{Kim:2002uw}.  The mass distribution is described in
detail in Ref.~\cite{Belczynski:2002}.  Of the injections performed, 76\% have
an associated coincident trigger in the LIGO and TAMA300 detectors. The
majority of the injections not detected have an effective distance at the
TAMA300 site greater than TAMA300's range during DT8.  However, there were
also a few injections which were very poorly oriented for the LIGO detectors,
and hence have a large effective distance, making them unobservable to LIGO.
Finally, several injections produce triggers in both the LIGO and TAMA300
detectors but these fail our coincidence requirements.  The SNRs of these
triggers are close to threshold in TAMA300 and the injection parameters, in
particular the chirp mass, are recovered poorly. In Figure
\ref{fig:shift_inj_trigs} we plot the coincident triggers associated with
injections superimposed on those from the time-shift analysis.  This shows
that triggers from the found injections are well separated from the accidental
coincidences found in the time-shift analysis.   
 
\begin{figure} 
\begin{center} 
\includegraphics[width=\linewidth]{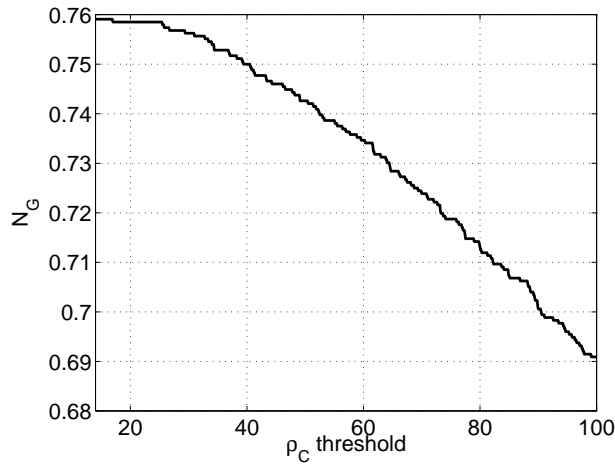} 
\end{center} 
\caption{The efficiency of the LIGO--TAMA300 joint analysis to simulated 
galactic inspiral events.  The number of galaxies ($N_{G}$) to which the search 
is sensitive is plotted as a function of the threshold on the combined statistic 
$\rho_{\mbox{\rm\tiny C}} \ ( = \rho_{\mbox{\rm\tiny LIGO}} + 
\rho_{\mbox{\rm\tiny TAMA}})$.} 
\label{fig:inj_eff} 
\end{figure}

In Figure \ref{fig:inj_eff},  we plot the sensitivity of the search to
injected Milky Way signals.  For consistency with previous searches
\cite{LIGOS2iul} we use $N_{G}$ to represent the number of galaxies to which
the search is sensitive.  For this search, $N_{G}$ is equivalent to the
fraction of Milky Way signals we are sensitive to.  The figure shows the
number of galaxies the search is sensitive to as a function of the threshold
on the combined statistic given in Eq. (\ref{eq:rho_combined}).  Thus, at
threshold we are sensitive to a little more than three quarters of candidate
sources in the galaxy.  The efficiency curve is used later in determining the
upper limit and associated systematic errors.

\begin{figure} 
\begin{center} 
\includegraphics[width=\linewidth]{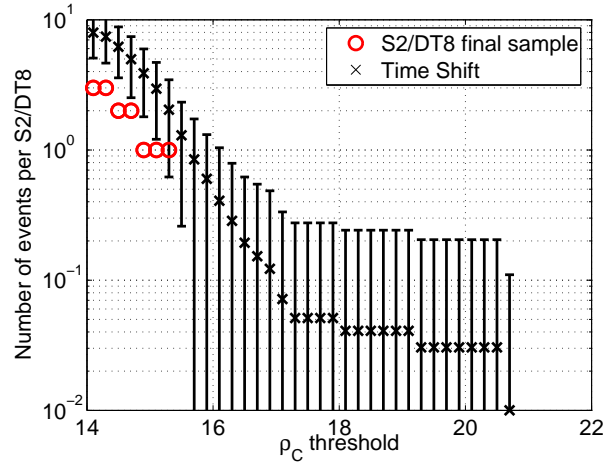} 
\end{center} 
\caption{The triggers from the analysis of the full LIGO--TAMA300 data set. 
The $\times$ represent the expected background number of triggers at or above 
a given combined SNR $\rho_{\mbox{\rm\tiny C}}$ based on the 100 time-shifts 
performed.  The bars indicate the standard deviation of the number of events, 
calculated from the time-shift results. 
The triggers from the final S2/DT8 data set are shown as $\circ$.} 
\label{fig:shift_zero_dist} 
\end{figure}

We analyze the S2/DT8 data using the pipeline described.  The cumulative
distribution of $\rho_{\mbox{\rm\tiny C}}$ of the coincident triggers is shown
in Figure~\ref{fig:shift_zero_dist}.  On this plot, the expected number of
triggers obtained from the time-shift analysis is shown, as well as the
standard deviation of the number of triggers obtained in the time-shifts. The
results of the analysis of the full data are overlayed on top of this. It is
clear from the figure that the distribution of coincident triggers is
consistent with the background estimated from time-shifts.  There are no
triggers with combined SNR greater than $\rho_{\rm max} = 15.3$.  Therefore,
we conclude that there is no evidence for gravitational wave signals in the
LIGO--TAMA300 S2/DT8 data set.   
 
Given the set of triggers displayed in Figure~\ref{fig:shift_zero_dist} we can 
obtain an upper limit on the rate of binary neutron star coalescences per year 
per Milky Way Equivalent Galaxy (MWEG).  (Although this search is only 
sensitive to galactic inspiral events, we maintain the standard ``MWEG'' 
\cite{LIGOS2iul} for describing the upper limit).  We use the loudest event 
statistic~\cite{loudestGWDAW03}, which makes use of the detection efficiency at 
the combined SNR of the loudest event in order to construct the upper limit. 
The 90\% confidence frequentist upper limit is given by 
\begin{equation} \mathcal{R}_{90\%} = \frac{2.303 + \ln P_b}{T 
N_G(\rho_{\rm\tiny max})} \, . 
\end{equation} 
In the above, $T$ is the observation time of 584 hours, $P_{b}$ is the
probability that all background triggers have a SNR less than $\rho_{\rm\tiny
max}$, and $N_{G}$ is the number of MWEGs the search is sensitive to at the
combined SNR of the loudest event $\rho_{\rm\tiny max}$.  $N_{G}$ is
determined from Figure \ref{fig:inj_eff} to be $0.76 \, \mathrm{MWEG}$ for
$\rho_{\rm\tiny max} = 15.3$.  Although the time-shift analysis provides us
with an estimate of $P_{b} = 0.2$, we note that it is difficult to establish a
systematic error associated with this estimate, and therefore take the
conservative choice of setting $P_{b} = 1$.  From these numbers, we obtain an
upper limit of $\mathcal{R}_{90\%} = 45 \, \mathrm{y}^{-1}\,
\mathrm{MWEG}^{-1}$. 
 
The possible systematics which arise in a search for binary neutron stars are 
described in some detail in Ref.~\cite{LIGOS2iul}, and we will follow the 
analysis presented there to calculate the systematic errors for the above 
result.  The most significant effects are due to the possible calibration 
inaccuracies of the detectors, the finite number of Monte Carlo injections 
performed,  and the mismatch between our search templates and the actual 
waveform.  We must also evaluate the systematic errors associated with the 
chosen astrophysical model of potential sources within the galaxy.  All 
systematic effects in the analysis pipeline (such as less than perfect 
coverage of the template bank) are taken into account in the Monte Carlo 
estimation of the detection efficiency.   
 
This search was sensitive to most, but not all, signals from the Milky Way.
Thus, the specific model of the source distribution within the galaxy will
affect the upper limit.  The majority of the mass in the galaxy, and hence the
potential sources, is concentrated near the galactic center.  Therefore, our
efficiency will be most affected by changing the distance from the sun to the
center of the galaxy in the model.  In this search, the sun's galactocentric
distance is assumed to be $8.5 \, \mathrm{kpc}$.  Varying this distance
between $7$ and $10 \, \mathrm{kpc}$ leads to a change in efficiency of $0.04
\, \mathrm{MWEG}$.  Different models for NS-NS formation can lead to
variations in the NS mass distribution.  Based on simulations with a 50\%
reduction in the number of binary systems with masses in the range $1.5
M_{\odot} < m_{1}, m_{2} < 3.0 M_{\odot}$, we can estimate the variation in
$N_{G}$ to be $0.01 \, \mathrm{MWEG}$ 
 
Any calibration inaccuracy in TAMA300 could have a significant effect upon our 
efficiency.  This is clear from Figure~\ref{fig:shift_inj_trigs} which shows a 
significant number of injections found in TAMA300 close to threshold.  
Two effects contribute to this calibration error: an overall normalization
error (associated with the magnetic actuation strength uncertainty and its
effect on calibration), and uncertainty in the frequency-dependent response.
The error in the normalization is of order 5\%, but the long-term drift is
unknown, so we conservatively use 10\% in this paper.  The frequency-dependent
error was estimated and shown to be $\ll 10\%$, so it is subsumed into the
overall 10\% error on the SNR of the triggers.  This calibration uncertainty
leads to a $0.02 \, \mathrm{MWEG}$ effect on our efficiency.  The majority of
injections are observed well above threshold in the LIGO detectors, and
consequently the calibration uncertainty of $8.5\%$ in L1 and $4.5\%$ in H1/H2
results in a smaller uncertainty in the efficiency of $< 0.01 \,
\mathrm{MWEG}$.  The error in the efficiency measurement due to the finite
number of injections performed is $0.01 \, \mathrm{MWEG}$.  Differences
between the theoretical waveforms used in matched filtering the data and the
real waveforms would decrease the efficiency of our search.  Allowing for a
$10\%$ loss in SNR due to inaccuracies in the model
waveform~\cite{Apostolatos:1995, Droz:1997, Droz:1998} leads to a $+0/-0.02 \,
\mathrm{MWEG}$ effect on the efficiency.  Combining these effects, we obtain
an efficiency of $N_{G} = 0.76^{+0.05}_{-0.06}$. Taking the downward excursion
on $N_{G}$, we obtain a conservative upper limit of 
\begin{equation}  
  \mathcal{R}_{90\%} = \upperlimit \, \mathrm{y}^{-1} \, 
  \mathrm{MWEG}^{-1} \, .   
\end{equation} 

This rate is substantially higher than the predicted astrophysical rate of
$8.3 \times 10^{-5}  \mathrm{y}^{-1} \, \mathrm{MWEG}^{-1}$
\cite{Kalogera:2003tn}.  However, the rate limit obtained in this paper is
comparable with the rate limit of $47 \, \mathrm{y}^{-1} \,
\mathrm{MWEG}^{-1}$ obtained from the LIGO-only S2 search~\cite{LIGOS2iul},
which was performed on a complementary data set.  Since these searches were
performed on independent data sets, if astrophysically relevant, these upper
limits could then be combined to produce the best possible limit.  The fact
that the LIGO S2 and LIGO-TAMA300 S2/DT8 limits are so similar demonstrates that
the overall sensitivities of the two searches are very nearly equal.  This is
achieved despite the fact that the TAMA300 detector was less sensitive than LIGO
during S2/DT8.  The high duty cycle of TAMA300 (over 80\%) compensates for the
reduced sensitivity, and leads to a similar overall result.  

In this paper, we have presented the methods and results from the first
multi-collaboration, network search for gravitational waves from inspiralling
binary systems using large scale interferometers.  The search was performed
using a trigger exchange method, requiring coincidence in both the end time
and chirp mass of triggers between instruments.  Using this method, we have
performed all necessary steps of the analysis, including time-shifts, signal
injections and the calculation of the upper limit.  The joint, coincidence
search presented here is a natural first step in any network analysis.   The
methods developed during this search will be applied in future network
searches.  Indeed, the experience gained during this joint search is being
used in subsequent LIGO Scientific Collaboration searches of LIGO and GEO
data.  Furthermore, a trigger exchange coincidence analysis is being developed
as the first stage of a future joint LIGO--Virgo analysis.  The optimal
network search would likely involve a fully coherent analysis \cite{ref:Pai}
of the detectors' data streams around the times of coincident events.  A
coherent followup to the coincidence method presented in this paper would be
included in future analyses of LIGO and TAMA300 data with improved sensitivity,
or in joint analyses of the planned second generation detectors such as
advanced LIGO \cite{advLIGO} in U.S. and LCGT \cite{LCGT} in Japan.

The authors gratefully acknowledge the support of the United States National 
Science Foundation for the construction and operation of the LIGO Laboratory 
and the Particle Physics and Astronomy Research Council of the United Kingdom, 
the Max-Planck-Society and the State of Niedersachsen/Germany for support of 
the construction and operation of the GEO600 detector. The authors also 
gratefully acknowledge the support of the research by these agencies and by the 
Australian Research Council, the Natural Sciences and Engineering Research 
Council of Canada, the Council of Scientific and Industrial Research of India, 
the Department of Science and Technology of India, the Spanish Ministerio de 
Educacion y Ciencia, the John Simon Guggenheim Foundation, the Leverhulme 
Trust, the David and Lucile Packard Foundation, the Research Corporation, and 
the Alfred P. Sloan Foundation.  TAMA research is supported by a Grant-in-Aid 
for Scientific Research on Priority Areas (415) of the Ministry of Education, 
Culture, Sports, Science and Technology of Japan. This work was also supported 
in part by JSPS Grant-in-Aid for Scientific Research Nos.~14047214 and 
12640269.

\end{document}